\documentclass[10pt]{article}
\usepackage[a4paper, total={7in, 9in}]{geometry}
\usepackage[utf8]{inputenc}
\usepackage{graphicx}
\usepackage[dvipsnames]{xcolor}
\usepackage{amsmath,amssymb}
\usepackage{soul}
\usepackage{multirow}
\usepackage{authblk}
\usepackage{hyperref}
\newcommand{\link}[1]{\href{#1}{#1}}

\begin{document}

\title{A statistical analysis of death rates in Italy for the years 2015--2020 and a  comparison with the casualties reported for the COVID-19 pandemic.}

\author[1,2]{Gianluca Bonifazi}
\author[3,4]{Luca Lista}
\author[5,*]{Dario Menasce}
\author[6]{Mauro Mezzetto}
\author[2]{Alberto Oliva}
\author[5]{Daniele Pedrini}
\author[2]{Roberto Spighi}
\author[7,2]{Antonio Zoccoli}
\affil[1]{\raggedright\normalsize Universit\`a Politecnica delle Marche}
\affil[2]{\raggedright\normalsize INFN Sezione di Bologna}
\affil[3]{\raggedright\normalsize Universit\`a degli Studi di Napoli Federico II}
\affil[4]{\raggedright\normalsize INFN Sezione di Napoli}
\affil[5]{\raggedright\normalsize INFN Sezione di Milano Bicocca}
\affil[6]{\raggedright\normalsize INFN Sezione di Padova}
\affil[7]{\raggedright\normalsize Alma Mater Studiorum Universit\`a di Bologna}
\affil[*]{\raggedright\normalsize \textbf{Corresponding author}, e-mail: {\tt dario.menasce@mib.infn.it}}
\date{}

\maketitle

\begin{abstract}
We analyze the data about casualties in Italy in the period 01/01/2015 to 30/09/2020 released by the Italian National Institute of Statistics (ISTAT). Aim of this article is the description of a statistically robust methodology to extract quantitative values for the seasonal excesses of deaths featured by the data, accompanying them
with correct estimates of the relative uncertainties. We will describe the advantages of the method adopted with respect to others listed in literature. The data exhibit a clear sinusoidal behavior, whose fit allows for a robust subtraction of the baseline trend of casualties in Italy,  with a surplus of mortality in correspondence to the flu epidemics in winter and to the hottest periods in summer. 
 The overall quality of the fit to the data turns out to be very good, an indication of the validity of the chosen model. We discuss the trend of casualties in Italy by different classes of ages and for the different genders.
We finally compare the data-subtracted casualties, as reported by ISTAT, with those reported by the Italian Department for Civil Protection (DPC) relative to the deaths directly attributed to COVID-19, and we point out the differences in the two samples, collected under different assumptions.
\end{abstract}

\section{Introduction}

Last year has seen an enhanced global attention concerning the death rates in various countries, a fact due to the outbreak of the COVID-19 pandemics at the beginning of 2020 and the ensuing alarm it generated worldwide. A first visual inspection of the historical data archives\cite{ISTAT-Data}, publicly made available by the Italian Istituto Nazionale di Statistica (ISTAT)\cite{ISTAT}, shows a periodic variation of the death rate depending upon seasons, well represented by a stable and regular sinusoid. Superimposed  to the sinusoid trend there may be additional death excesses, most likely due to seasonal diseases like influenza in winter or to very intense heat waves in the summer \cite{excessDeaths-1,excessDeaths-2}. 

The approach adopted here for an estimate of the seasonal excess of deaths is an interpolation of the data with a fit function exhibiting an \emph{ad hoc} modeling of the main features of the curve. While the excess peaks are symmetric in shape, the peak in coincidence with the COVID-19 pandemics is asymmetric and more pronounced. We fit the former with a Gaussian function and the latter  with a Gompertz function, in order to quantify number of casualties, the duration and the position of all causes of excess deaths. Focus of this paper is the method to compute the number of deaths from the data rather than a discussion about the particular functional model chosen for this task or an interpretation of the outcomes of such an evaluation. A least-squares fit of a single function, encompassing both the background (the periodical seasonal variation of deaths) and the specific additional excesses above this background, allows for a very robust evaluation of the latter, both for the numerical values and for the relative uncertainties. We present results of this method applied to the data provided by ISTAT in the period 2015--2020. A comparison is then carried out with a different data sample\cite{ProtezioneCivile-Data}, provided by the Dipartimento della Protezione Civile (DPC) \cite{ProtezioneCivile}, which provides counts for deaths directly attributed to the COVID-19 disease.

\section{The data sample}
This study is based on publicly available data provided by ISTAT \cite{ISTAT-Data} as time series of recorded deaths by the National Registry Office. The data, collected from all the 7903 districts located in the 20 Italian regions, covers the period from January $1^{\mathrm{st}}$, 2015 to September $30^{\mathrm{th}}$, 2020. We have collected these data in histograms where each bin contains the number of deaths for a single day.

\begin{figure}[htbp]
    \centering
    \includegraphics[width=0.99\textwidth]{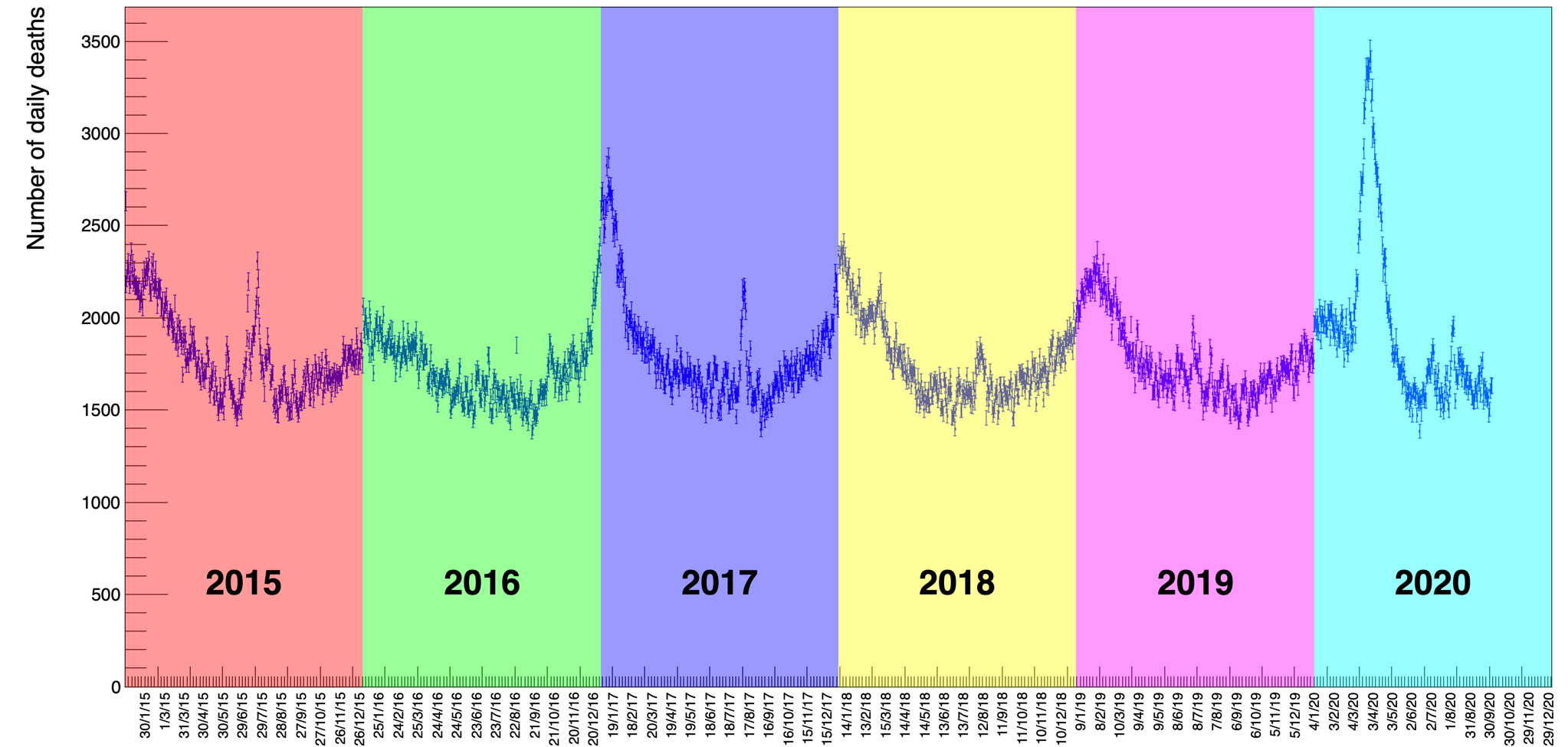}
    \caption{The distribution of deaths collected by ISTAT \cite{ISTAT-Data} from 7903 districts in Italy between January $1^{\mathrm{st}}$ 2015 to September $30^{\mathrm{th}}$ 2020. These data include both genders and all ages. An annual modulation of the counts is evident with maxima corresponding to winter seasons and minima to summer.}
    \label{fig:figYears}
\end{figure} 

The data are collected by gender, age and location  for each individual death. Figure~\ref{fig:figYears} shows the number of deaths in all the categories in the considered period. 
What is already striking by a simple visual inspection of the distribution is a periodic seasonal variation that behaves approximately like a  sinusoidal wave of constant amplitude on top of an equally constant offset value\footnote{In the following we will discuss how we established that there is no significant slope of the average value of this wave.}. This feature remains partly confirmed also by disentangling the data using age as a selection criteria. 
\begin{figure}[htbp]
    \centering
    \includegraphics[width=0.99\textwidth]{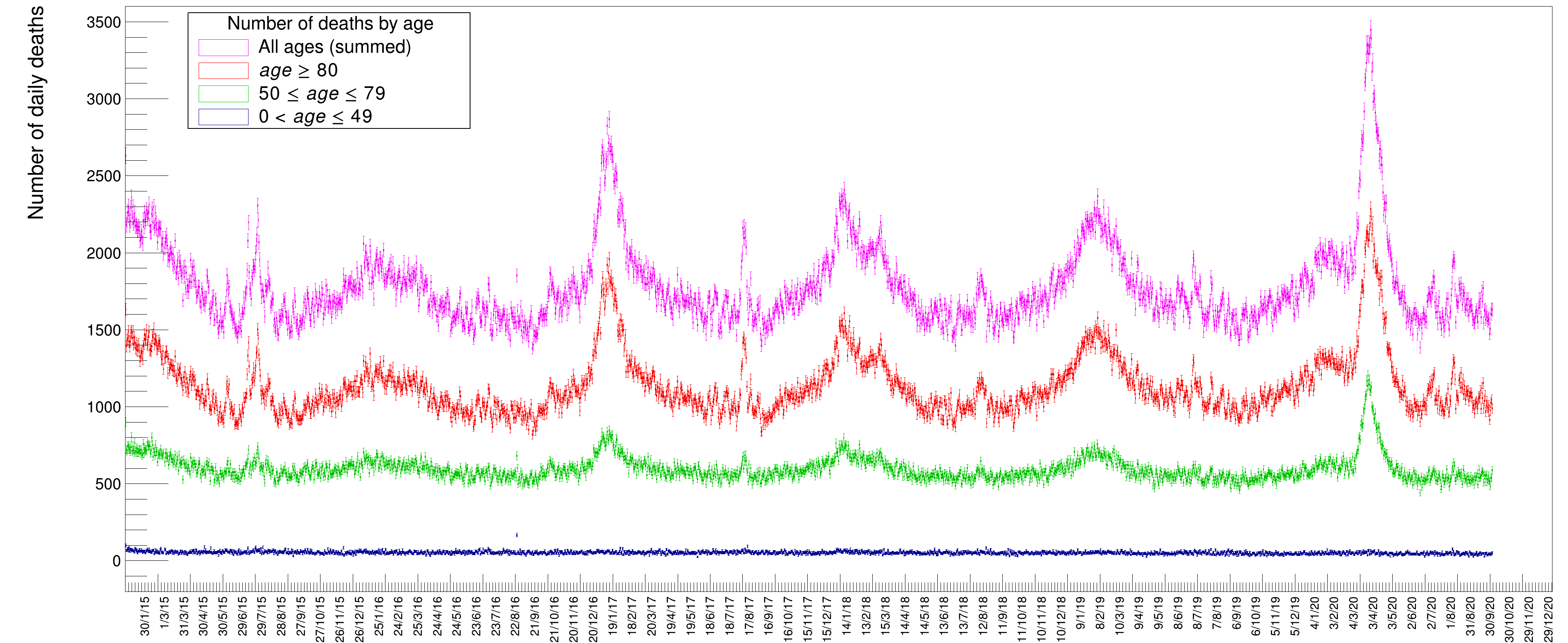}
    \caption{Distribution of number of deaths along six years for specific age intervals.}
    \label{fig:figDisentangle}
\end{figure} 
Fig.~\ref{fig:figDisentangle} shows the distribution of deaths for people in three different age classes: in blue those below 50 years, in green those in range 50 to 79; in red those above 80 and in magenta the sum of all these three classes. It is evident from these distributions that people with an age below 50 die, to a good approximation, with a constant average probability in any given day of the year while those above that age tend to have a varying, periodic probability of death with a maximum in winter and a minimum in summer. The older the age, the larger the excess of death in particular periods of the year, appearing in the form of Gaussian-like excesses over the sinusoidal wave. Disentangling the data by gender, see Fig.~\ref{fig:figSexes}, there seems to be a slight prevalence of female deaths with respect to males, except for the COVID-19 peak, where the situation happens to be reversed. These are just raw values, though, not corrected to take into account the ratio between males and females in the Italian population. Later on, in this paper, we will quantify and appropriately weight these data.
\begin{figure}[htbp]
    \centering
    \includegraphics[width=0.99\textwidth]{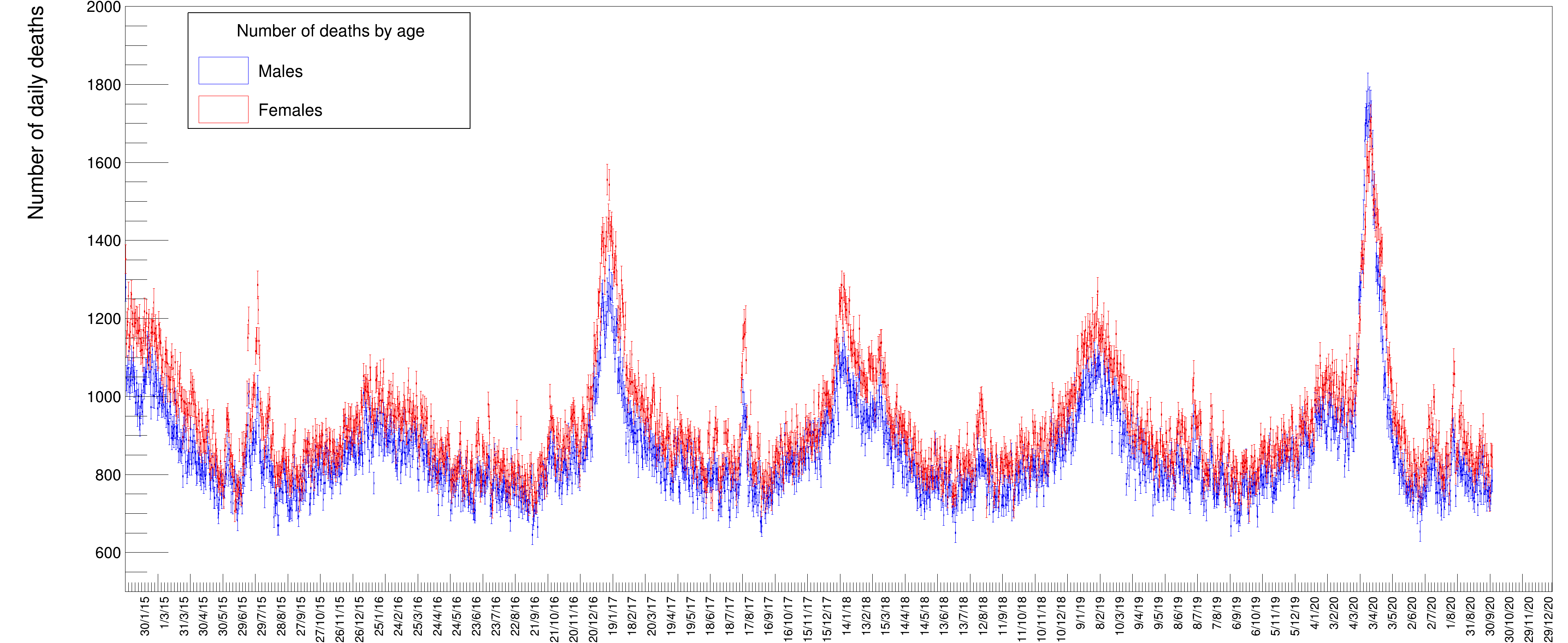}
    \caption{Distribution of number of deaths for males and females in the same time period of Fig~\ref{fig:figDisentangle}.}
    \label{fig:figSexes}
\end{figure} 
\hfill \break
\section{Methodology of the data analysis}

We perform a global fit of the data, where we simultaneously estimate the sinusoidal baseline of the distribution, the seasonal death excesses and the 2020 peak in correspondence of the COVID-19 pandemic. This method significantly differs from other methods often reported in literature\cite{OurWorldInData}. 

In particular we quote analyses \cite{HumanMortalityDatabase, WeeklyDeathStatistics, burden, temporalDynamics, excessUS, crossRegionalAnalysis, empiricalModel, outbreakMonitorig} in which the background is subtracted by computing the average number of counts in the same period of the past 5 years. In this way the excesses of seasonal pandemics, like flu, are expressed against the average counts of the same pandemics in the previous years and not in absolute terms. 

In \cite{Euromomo} the sinusoidal baseline is computed by fitting the data during periods of time where the excesses are not evident, like in spring or autumn. While this approach can be easily automated, it is subject to a certain degree of arbitrariness due to the specific choice of the periods included in the fit. 

A global fit to the time series has, instead, the merit of simultaneously using all the available data to shape both the excesses and the baseline, without any degree of arbitrariness. Furthermore the least-squares method provides a complete and fully correct covariance matrix that allows to compute the uncertainties involved in the final result. Eventually the goodness of the fit and the absence of biases can be quantified by the final $\chi^2$ of the interpolation and by the pulls distribution respectively.

We therefore used a $\chi^{2}$ fit to interpolate the data with an appropriate function meant to model the data in order to determine the value of the unknown parameters of the model along with their uncertainties. The actual minimization is carried out by the MINUIT~\cite{MINUIT} package, while the adopted statistical methodology is described in~\cite{James}.

The overall fit function has been defined as the sum of individual components in the following form:

\begin{equation}
  \label{fullFitFunction}
  F(t)=s(t)+{\sum_{i=1}^{k}g_{i}(t)}+\dot{G}(t)
\end{equation}

where $s(t)$, $g_{i}(t)$ end $\dot{G}(t)$ are defined and described below. 

The $s(t)$ function is meant to model the wave-like variation of deaths with seasons, the 
$g_{i}(t)$ function describes the excess peaks visible above the wave and $\dot{G}(t)$ represents the rightmost excess peak (spring 2020), which, unlike the others, is asymmetrical.
The index $i$ runs from 1 up to $k$, the number of excess peaks featured by the data distribution except the last one on the far right ($k=13$ peaks in this particular case).

The general wave-like behavior of the data is modeled by a sinusoidal function of the form:

\begin{equation}
  \label{waveFunction}
  s(t)=c(t) + a \sin\left(\frac{2\pi t}{T}+\varphi\right)\, 
\end{equation}

where $t$ is the day number starting from $t_0=1/1/2015$. The parameter $c(t)$ represents the slowly-varying offset from zero deaths, $a$ the amplitude of the oscillation, $T$ is the period of variation (the time delay between consecutive maxima) and finally $\varphi$ the phase.
We tried to model $c(t)$ allowing for a linear dependence on $t$, as $c(t)=c_0+c_1t$, but the fit determines a slope $c_1$ compatible to zero within uncertainty. We therefore decided to maintain the $c$ term constant and independent of time in the final fit.

Each individual excess above this $s(t)$ wave can be modeled by a Gaussian distribution of the canonical form:

\begin{equation}
 \label{gaussianFunction}
  g_i(t)=\frac{N_i}{\sqrt{2\pi}\sigma_{i}}e^{-(t-\mu_i)^2/2\sigma_i^2}
\end{equation}

The choice of a Gaussian function here is only justified by being the simplest symmetrical function to describe these excesses representing, at the same time the distribution of a random variable.
Modeling the excess peaks in the described way has the advantage that the individual $g_{i}$ fit parameters correspond to a Gaussian with the background contribution already taken into account in the overall fit model. The $N_i$ parameter of each Gaussian, corresponds to the number of excess deaths with respect to the wave-like background, whose values are also determined optimally by the fit itself. An advantage of this approach is that in the case of adjacent, overlapping Gaussians (as can be seen in Fig.~\ref{fig:figAll} in the case of the $g_6$ end $g_7$ peaks but also the $g_{11}$ and the big peak on the far right of the distribution), each individual area is computed correctly by taking into account the nearby contributing ones.

While the excess peaks look highly symmetrical around their maximum and can thus be reasonably well modeled with Gaussians, as described before, the peak of the spring 2020, associated with the COVID-19 pandemic, is clearly asymmetric.
We have tried several possible parametrizations for that distribution, such as bifurcated Gaussians with a common peak, generalized logistics, or else, to reflect the asymmetry, but in the end we resolved to adopt the derivative of a Gompertz function\cite{GompertzArticle, Gompertz} simply because it is customarily adopted by epidemiologists to describe epidemic evolution's over time and we therefore considered it more suitable to our purpose.

A Gompertz function is parametrized in the following way:

\begin{equation}
  G(t)=N_G\,e^{-b\,e^{-ht}}\,
  \label{eq:gompertz}
\end{equation}

Equation~\ref{eq:gompertz} represents a cumulative distribution. Since our data represent instead daily counts, we used its derivative, given by:

\begin{equation}
  \dot{G}(t)=\frac{\mathrm{d} G(t)}{\mathrm{d} t} =  N_G\,bh\,e^{-b\, e^{-ht}}\,e^{-ht}\,
\end{equation}
where the parameter $N_G$ is the value of the integral of this function.
It is worthwhile to note that a global fit can correctly take into account contributions from partially overlapping peaks, like $g6$ and $g7$ or $g11$ and $\dot{G}$ in Figure~\ref{fig:figAll}, something that no other method can accomplish correctly.

\section{Results and discussion}

\begin{figure}[htbp]
    \centering
    \includegraphics[width=0.99\textwidth, height=0.28\paperheight]{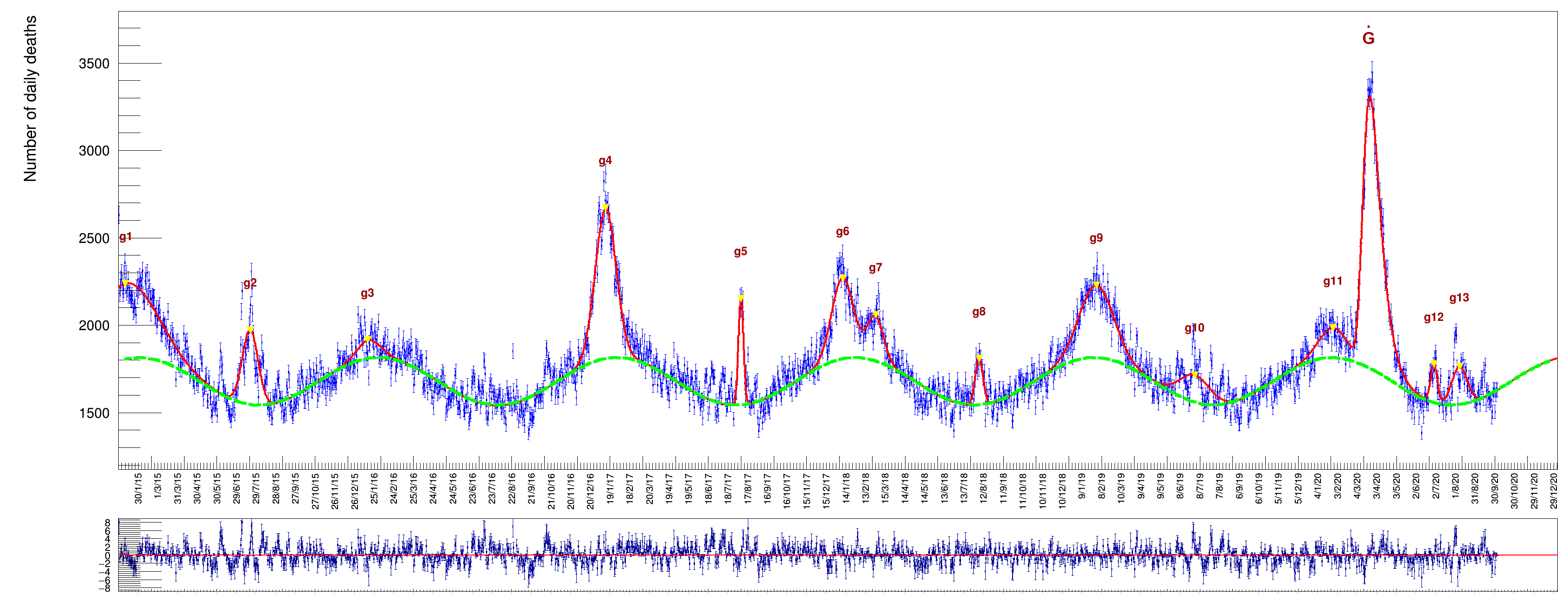}
    \caption{The whole data sample with a superimposed fit function obtained as specified in the text. The plot at the bottom shows the \emph{pulls} (a quantity defined later on in the text) of the fit: their mean value, being compatible with zero and the absence of remarkable localized deviations from this value along the whole time series, is a testimony of the appropriate choice of the particular model adopted. Numerical values for the integral of each individual Gaussian are provided in Table~\ref{table:allGPars}. The $g1$ to $g13$ labels indicate the 13 Gaussians  introduced to describe the data (see eq.~\ref{gaussianFunction}).}
 
    \label{fig:figAll}
\end{figure}

In Figure~\ref{fig:figAll} and Tables~\ref{table:allGPars} to~\ref{table:2} we report the
results of a fit to the whole data sample, comprising both genders for all ages in the six years from 2015 to 2020. The column labeled `$\mu_i$', in Table~\ref{table:allGPars}, indicates the day when the maximum of an excess has been reached while those labeled `$\mu_i\pm2\sigma_i$' indicate, respectively, the day of onset and demise from the average background, a time interval in which occur 95\% of the death cases (expressed with calendar dates). The column labeled `\emph{Duration}' is the time difference between onset and demise (namely $4\sigma$, expressed as number of days). 

\begin{table}[htbp]
\centering
\begin{tabular}{|| c | c | c | c | c | c | c | c ||} 
 \hline
 $g_i$ & \emph{$N_{i}$} & \emph{$\mu_{i}$} (days) & \emph{$\sigma_{i}$} (days) & $\mu_i-2\sigma_i$ (date) & $\mu_i$ (date) & $\mu_i+2\sigma_i$ (date) & \emph{Duration} (days)\\ [0.5ex] 
 \hline\hline
1  & 50\,706 $\pm$ 3\,092 &   10.9 $\pm$ 3.4 & 47.1 $\pm$ 2.3 &  9/10/2014 & 10/1/2015 & 15/4/2015 & 189   \\
2  & 13\,005 $\pm$  361 &  201.1 $\pm$ 0.4 & 12.0 $\pm$ 0.4 & 25/ 6/2015 & 20/7/2015 & 13/8/2015 & 49    \\
3  &  4\,455 $\pm$  527 &  381.6 $\pm$ 2.0 & 17.7 $\pm$ 2.1 & 12/12/2015 & 16/1/2016 & 21/2/2016 & 71    \\
4  & 34\,015 $\pm$  534 &  743.5 $\pm$ 0.2 & 15.9 $\pm$ 0.3 & 11/12/2016 & 12/1/2017 & 13/2/2017 & 64    \\
5  &  5\,959 $\pm$  210 &  950.1 $\pm$ 0.1 &  3.9 $\pm$ 0.2 & 30/7/2017 &  7/8/2017 & 14/8/2017 & 15    \\
6  & 19\,120 $\pm$  704 & 1104.9 $\pm$ 0.6 & 16.5 $\pm$ 0.8 &  6/12/2017 &  8/1/2018 & 11/2/2018 & 67    \\
7  &  7\,862 $\pm$  616 & 1155.0 $\pm$ 1.0 & 12.5 $\pm$ 1.2 &  3/2/2018 & 28/2/2018 & 24/3/2018 & 49    \\
8  &  4\,084 $\pm$  256 & 1313.3 $\pm$ 0.4 &  6.0 $\pm$ 0.5 & 24/7/2018 &  5/8/2018 & 17/8/2018 & 24    \\
9  & 26\,850 $\pm$  685 & 1492.5 $\pm$ 0.6 & 26.1 $\pm$ 0.6 & 10/12/2018 & 31/1/2019 & 24/3/2019 & 104   \\
10 &  9\,299 $\pm$  504 & 1642.3 $\pm$ 1.4 & 23.2 $\pm$ 1.3 & 15/5/2019 & 30/6/2019 & 15/8/2019 & 92    \\
11 &  9\,020 $\pm$  613 & 1853.3 $\pm$ 1.7 & 21.8 $\pm$ 1.5 & 14/12/2019 & 27/1/2020 & 10/3/2020 & 87    \\
12 &  2\,841 $\pm$  217 & 2007.5 $\pm$ 0.4 &  5.1 $\pm$ 0.4 & 19/6/2020 & 29/6/2020 &  9/7/2020 & 20    \\
13 &  6\,546 $\pm$  362 & 2046.3 $\pm$ 0.7 & 11.9 $\pm$ 0.8 & 14/7/2020 &  7/8/2020 & 31/8/2020 & 48    \\
 \hline
\end{tabular}
\caption{Results of the fit for individual parameters (and their associated error) for each Gaussian, as modeled by equation number \ref{gaussianFunction}. The columns header indicates the Gaussian number ($g_{i}$), the yield (its area, $N_{i}$), its peak position ($\mu_{i}$), the width (the one standard-deviation duration expressed in number of days, $\sigma_{i}$) and the duration within $4\sigma$ (the difference between the values of column 7 and 5, also expressed as number of days).}
\label{table:allGPars}
\end{table}

\begin{table}[htbp]
\centering
\begin{tabular}{|| c | c | c | c ||} 
 \hline
 \emph{$c$} & \emph{$a$} & \emph{$T$} (days) & \emph{$\varphi$} (rad)\\ [0.5ex] 
 \hline\hline
1678 $\pm$ 1.5 & 139.4 $\pm$ 2.593 & 364 $\pm$ 0.4 & -5.27 $\pm$ 0.02 \\ \hline
\end{tabular}
\caption{Results of the fit to the whole data set (no selection applied) for the baseline sinusoidal wave, as modeled by eq.~\ref{waveFunction}. 
The columns header indicates the average value of the sinusoid $ C$, the amplitude, $a$, the period, $T$ and the phase, $\phi$ as further explained in the text. }
\label{table:4}
\end{table}

The \emph{pulls}, $p_i$, are defined as:
\begin{equation}
    p_i = \frac{d_i - F(t_i)}{\epsilon_i}
\end{equation}
where  $d_{i}$ is the number of death counts in a given day $i$
and $\epsilon_i$ the corresponding amount of statistical fluctuation.
The data, being outcomes of counting values, are assumed to follow a Poisson distribution, hence $\epsilon_{i}=\sqrt{d_{i}}$.

The $\chi^{2}/{n_{\mathrm{DOF}}}$ of the fit turns out to be 3.271. 

\begin{figure}[htbp]
    \centering
    \includegraphics[width=0.5\textwidth]{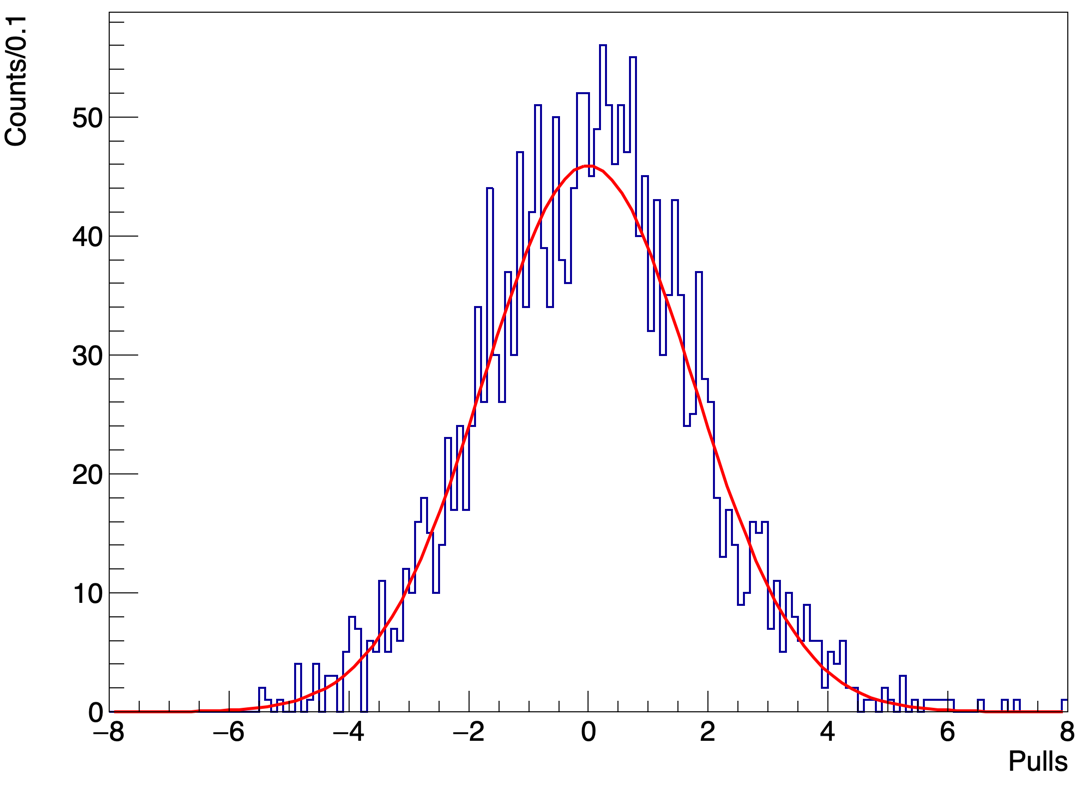}
    \caption{Distribution of the pulls (depicted as a time series in the bottom plot of Fig.~\ref{fig:figAll}) fitted with a Gaussian function. The peak of the Gaussian is at $\mu=-0.01\pm0.04$ (therefore compatible with zero) while the width id given by $\sigma=1.75\pm0.04$.}
    \label{fig:figPulls}
\end{figure}

We report the distribution of the \emph{pulls} in Fig.~\ref{fig:figPulls} fitted with a Gaussian function. The mean value of the fit is $-0.01\pm0.04$, compatible with zero, while the standard deviation of the Gaussian fit turns out to be $1.75\pm0.04$, confirming the significant underestimate of the uncertainties. This deviation from unity, of about $75\%$, gives an approximate amount of the increase that could be applied to the data errors to make them compatible with Poissonian values.

The area of each Gaussian function $i$ is given by the fit parameter $N_i$ defined in eq.~\ref{gaussianFunction}, while the area of the Gompertz derivative is the fit parameter $N_G$ in eq.~\ref{eq:gompertz}.

\begin{table}[htbp]
\centering
\begin{tabular}{|| c | c | c | c | c ||} 
 \hline
\emph{Yield} & \emph{From} & \emph{Peak} & \emph{To} & \emph{Duration} (days)\\ [0.5ex] 
 \hline\hline
54\,387 $\pm$ 557 & 20/2/2020 & 24/3/2020 & 11/5/2020 & 81\\
\hline
\end{tabular}
\caption{Results of the fit to the whole data set (no filters applied) for the Gompertz derivative function. The meaning of the columns labeled \emph{From}, \emph{Peak} and \emph{To} is explained in the text.}
\label{table:2}
\end{table}
The \emph{width} of the Gompertz is computed from the first day in which the integral of the function exceeds 2.5\% of the total to the day in which the integral reaches 97.5\% of the total. These two days are reported in Table~\ref{table:2} under the columns labeled \emph{From} and \emph{To}.

The value of the period $T=364.0\pm 0.4$ days of the sinusoidal wave is compatible with a full year cycle within about three standard deviations.
The offset value $c=1678\pm1.5$ can be assumed to represent the average number of deaths per day (the overall vertical offset of the sinusoid with respect to the zero value). Finally, from the results of Table~\ref{table:4}, it turns out that the peak of the sinusoid (the maximum
number of deaths) falls on January $31^{st}$ of every year. 

These results highlight an interesting feature of the COVID-19 deaths excess. As already noted, almost every winter there is a surplus of deaths with respect to the baseline, with the notable exception of the years 2015--2016 (a period with a particularly balm winter\cite{climate}, with a relatively small value of 4\,455 excess of casualties). 
The peak in the spring of 2020, instead, shows characteristics markedly different from the winter excesses of previous year in terms of amplitude, width and day of the year when the maximum is reached. In the following we will mention the possible implications of these differences.

As far as we could investigate in literature, we didn't find any mention of usage of the interpolation methodology we indicate in this paper, whereas the most common approach adopted is a subtraction of the baseline from previous years.

\section{Age and gender mortality}

We have also disentangled the data by age and gender and fit the distributions in these different categories to obtain accurate numerical values. 
\hfill \break
\begin{figure}[htbp]
    \centering
    \includegraphics[width=0.99\textwidth, height=0.25\paperheight]{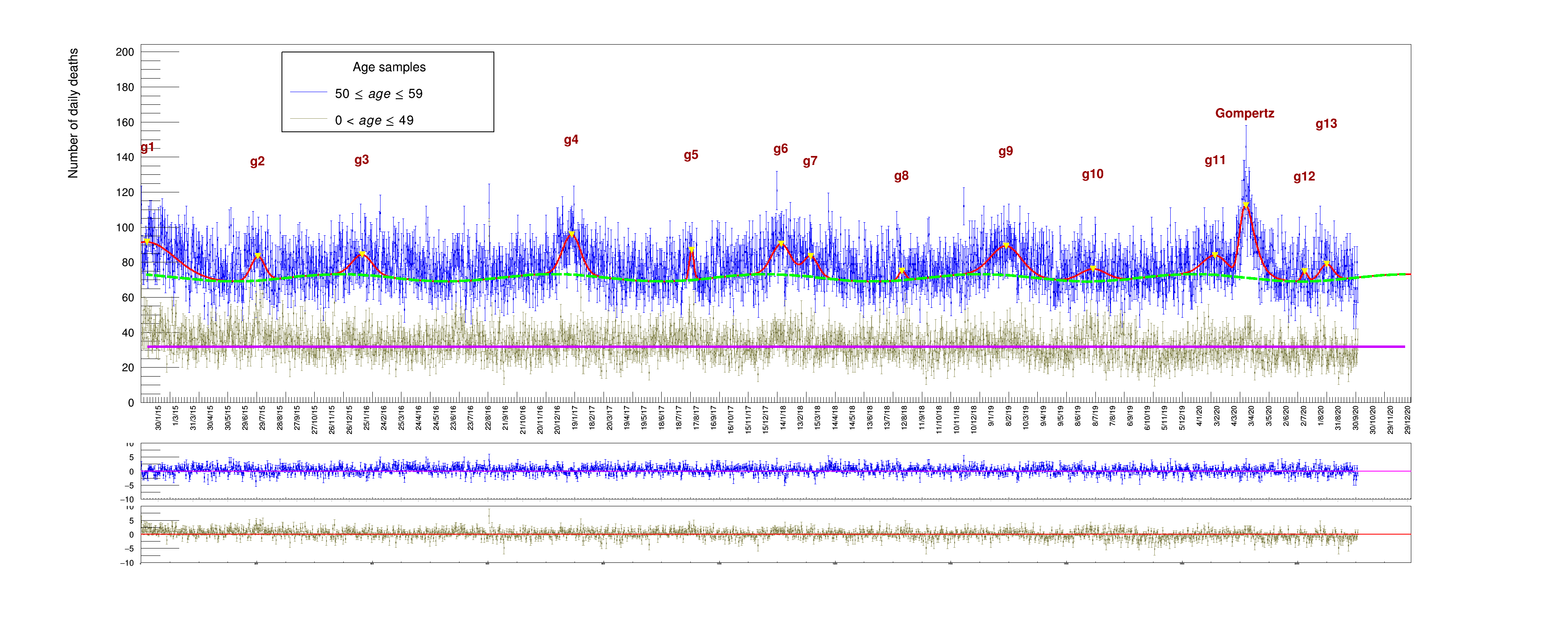}
    \caption{Casualties of people with age in the range $50\leq age \leq59$ with a superimposed fit based on function 6 (blue points) while gray points are the category $0<age\leq49$. Numerical results are listed in Table~\ref{table:5} and \ref{table:6}. The plots at the bottom show the \emph{pulls} of the two samples.}
    \label{fig:figLt60}
\end{figure} 

We start with a cumulative plot for all people aged between 50 (included) and 60 years (excluded) who died between 2015 and 2020, shown in Fig.~\ref{fig:figLt60}. This plot shows that the average value of daily deaths for people in this age range is about 70 casualties/day. In order to get a fit comparable with the one in Fig.~\ref{fig:figAll}, we are forced to adopt a somewhat more stringent fit strategy. 
\begin{table}[htbp]
\centering
\begin{tabular}{|| c | c ||} 
 \hline
 $g_i$ & \emph{Yield} \\ [0.5ex] 
 \hline\hline
  1 &       2252 $\pm$ 154 \\
  2 &        417 $\pm$  58 \\
  3 &        508 $\pm$  71 \\
  4 &        910 $\pm$  71 \\
  5 &        166 $\pm$  34 \\
  6 &        724 $\pm$  71 \\
  7 &        373 $\pm$  60 \\
  8 &         80 $\pm$  39 \\
  9 &       1105 $\pm$  90 \\
 10 &        413 $\pm$  78 \\
 11 &        603 $\pm$  79 \\
 12 &         71 $\pm$  37 \\
 13 &        276 $\pm$  57 \\
\hline
\end{tabular}
\caption{Results of the fit to the data set of people aged between 0 and 60 (excluded). These values correspond to the fit depicted in Fig.~\ref{fig:figLt60}
}
\label{table:5}
\end{table}
The wave parameter corresponding to the phase has been fixed to the value established for the full data sample (the other three are left free to float in the $s(t)$ function). This guarantees that \emph{maxima} are reached in the winter and \emph{minima} in the summer and no spurious time translation is introduced by the fit procedure when a local minima can eventually be found. Also the peak position and the width of the 13 Gaussians have been fixed to the values established by a fit to the whole data sample while the Gompertz parameters are all left free.  The corresponding fit results are listed in Tables \ref{table:5} and \ref{table:6}.

\begin{table}[htbp]
\centering
\begin{tabular}{|| c | c | c | c | c ||} 
 \hline
\emph{Yield} & \emph{From} & \emph{Peak} & \emph{To} & \emph{Duration} (days)\\ [0.5ex] 
 \hline\hline 
 1\,373 $\pm$  87 & 28/2/2020 & 24/3/2020 & 1/5/2020 & 63\\
 \hline
\end{tabular}
\caption{Results of the fit to the whole data set (no filters applied) for the Gompertz derivative function. These values correspond to the fit depicted in Fig.~\ref{fig:figLt60}.}
\label{table:6}
\end{table}

The picture shows two categories of age at the same time: those in the range $0-49$ (in gray) do not show any sign of seasonal variation around the mean value of $\sim32/\mathrm{day}$ (they were fit with a simple constant term). A sinusoidal variation begins to be noticeable only in the range $50-59$ (blue dots), along with the presence of the corresponding death excesses indicating a continuous increase in magnitude with age starting around 50.
The results are affected by larger uncertainties with respect to the full sample of Fig.~\ref{fig:figAll}, reflecting the smaller size of population in this range. 

\begin{figure}[htbp]
    \centering
    \includegraphics[width=0.99\textwidth]{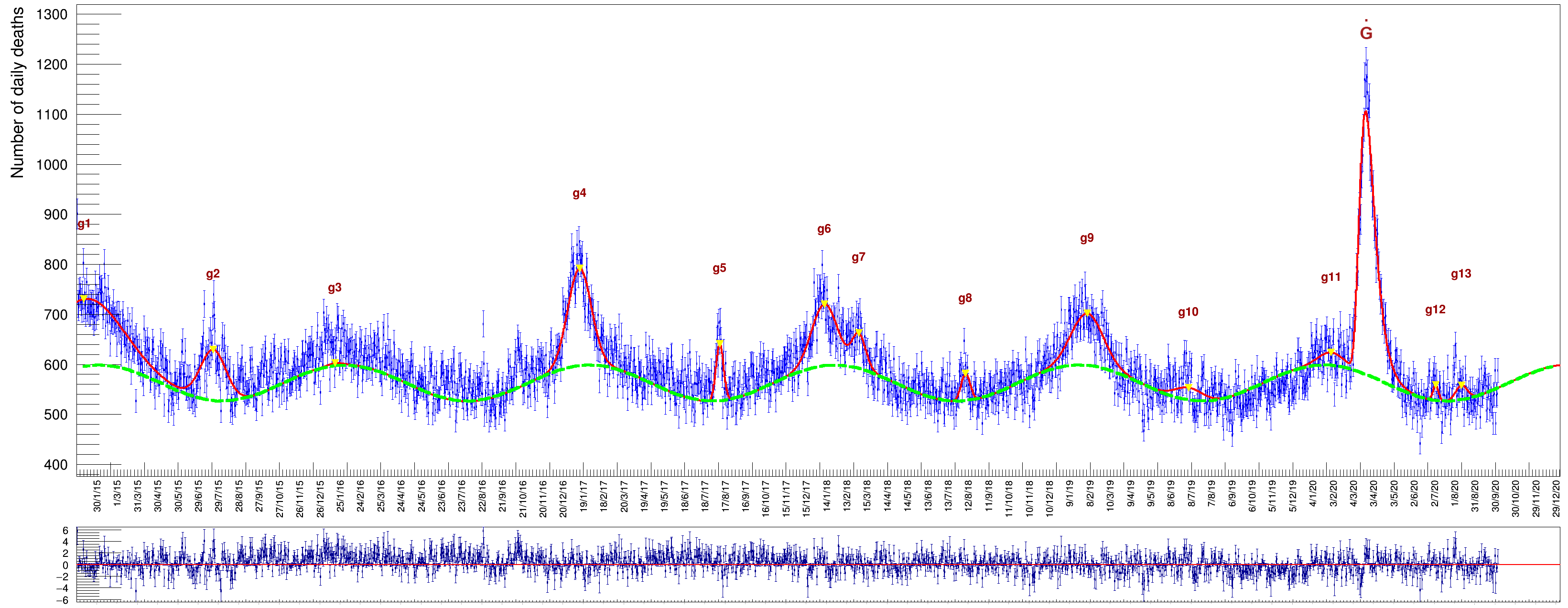}
    \caption{Casualties of people with $60 \leq \mathrm{age} <80$ (with a superimposed fit based on eq.\ref{fullFitFunction}). Numerical results are listed in Table~\ref{table:7} and \ref{table:8}}
    \label{fig:figGt60Lt80}
\end{figure}

The excess peaks and the sinusoid amplitude become more evident in a sample of even higher ages, namely $60 \leq \mathrm{age} < 80$. The average number of deaths in this category is much larger, due to an enhanced health fragility for people of progressively higher age, as seen in Fig.~\ref{fig:figGt60Lt80}.   

The fit is again pretty similar, in shape but not in amplitude of course, to the full sample shown in Fig.~\ref{fig:figAll}. The \emph{pulls} feature a mean value compatible with zero also in this case. The fit strategy is the same as the one described before for Fig.~\ref{fig:figLt60}. 
Values obtained in this case are listed in Tables \ref{table:7} and \ref{table:8}.
The average death rate in this category is $\sim72/day$.
\begin{table}[htbp]
\centering
\begin{tabular}{|| c | c ||} 
 \hline
 $g_i$ & \emph{Yield} \\ [0.5ex] 
 \hline\hline
  1 &      18\,225 $\pm$ 473 \\
  2 &       4\,916 $\pm$ 204 \\
  3 &          100 $\pm$   8 \\
  4 &       8\,432 $\pm$ 215 \\
  5 &       1\,591 $\pm$ 110 \\
  6 &       5\,851 $\pm$ 218 \\
  7 &       1\,704 $\pm$ 153 \\
  8 &          921 $\pm$ 117 \\
  9 &       6\,131 $\pm$ 239 \\
 10 &       1\,319 $\pm$ 200 \\
 11 &       1\,161 $\pm$ 203 \\
 12 &          320 $\pm$  89 \\
 13 &          657 $\pm$ 131 \\
 \hline
\end{tabular}
\caption{Results of the fit for the category of $50 \leq \mathrm{age} < 80$ . These values correspond to the fit depicted in Fig.~\ref{fig:figGt60Lt80}}
\label{table:7}
\end{table}
\begin{table}[htbp]
\centering
\begin{tabular}{|| c | c | c | c | c ||} 
 \hline
\emph{Yield} & \emph{From} & \emph{Peak} & \emph{To} & \emph{Duration} (days)\\ [0.5ex] 
 \hline\hline
15\,951 $\pm$ 242 & 25/2/2020 & 22/3/2020 & 25/4/2020 & 60\\
 \hline
\end{tabular}
\caption{Results of the fit for the category of $60 \leq \mathrm{age} < 80$  for the Gompertz derivative function. These values correspond to the fit depicted in Fig.~\ref{fig:figGt60Lt80}}
\label{table:8}
\end{table}
\begin{figure}[htbp]
    \centering
    \includegraphics[width=0.99\textwidth]{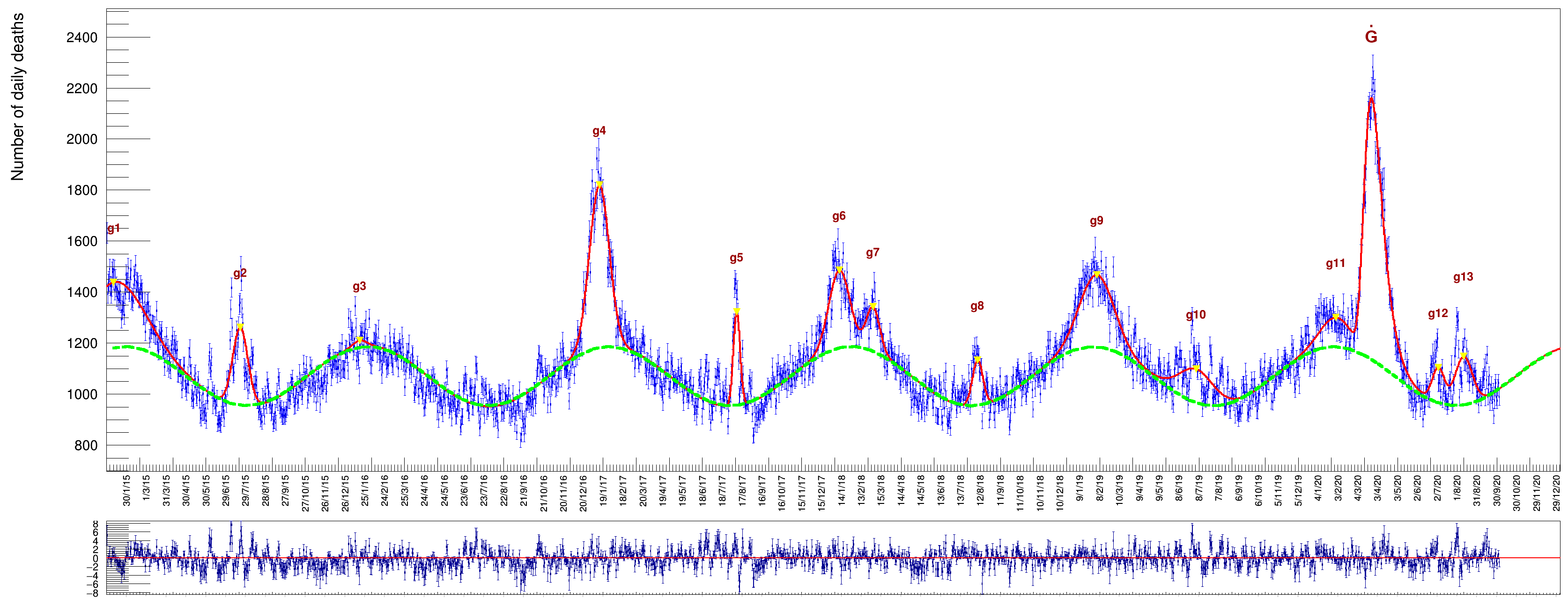}
    \caption{Casualties of people with $\mathrm{age} \geq 80$ (with a superimposed fit based on function 6). Numerical results are listed in Table~\ref{table:9} and \ref{table:10}}
    \label{fig:figGt80}
\end{figure} 
\begin{table}[htbp]
\centering
\begin{tabular}{|| c | c | c | c ||} 
 \hline
 $g_i$ & \emph{Yield} & \emph{Peak}& \emph{Duration} (days)\\ [0.5ex] 
 \hline\hline
  1 &      28\,324 $\pm$ 721 & 10/1/2015 & 176    \\
  2 &       8\,844 $\pm$ 254 & 21/7/2015 & 45     \\
  3 &          630 $\pm$ 220 & 17/1/2016 & 42     \\
  4 &      23\,732 $\pm$ 344 & 13/1/2017 & 60     \\
  5 &       4\,655 $\pm$ 179 &  9/8/2017 & 20     \\
  6 &      11\,978 $\pm$ 343 & 11/1/2018 & 63     \\
  7 &       4\,367 $\pm$ 267 &  3/3/2018 & 41     \\
  8 &       3\,607 $\pm$ 233 &  8/8/2018 & 33     \\
  9 &      18\,217 $\pm$ 425 &  4/2/2019 & 105    \\
 10 &       8\,981 $\pm$ 379 &  4/7/2019 & 107    \\
 11 &       6\,648 $\pm$ 389 & 31/1/2020 & 95     \\
 12 &       3\,208 $\pm$ 238 &  4/7/2020 & 37     \\
 13 &       6\,396 $\pm$ 270 & 11/8/2020 & 53     \\
\hline
\end{tabular}
\caption{Results of the fit to the data set of people aged over 80. These values correspond to the fit depicted in Fig.~\ref{fig:figGt80}}
\label{table:9}
\end{table}
\begin{table}[htbp]
\centering
\begin{tabular}{|| c | c | c | c | c ||} 
 \hline
\emph{Yield} & \emph{From} & \emph{Peak} & \emph{To} & \emph{Duration} (days)\\ [0.5ex] 
 \hline\hline
37\,357 $\pm$ 365 & 22/2/2020 & 25/3/2020 & 5/5/2020 & 73\\
 \hline
\end{tabular}
\caption{Results of the fit to the data set of people aged over 80 for the Gompertz derivative function. These values correspond to the fit depicted in Fig.~\ref{fig:figGt80}}
\label{table:10}
\end{table}
Increasing the age threshold further up, by collecting deaths of people aged $\geq 80$, we get a sample with very pronounced peaks, see Fig.~\ref{fig:figGt80}.
The average death rate in this last category reaches the high value of $\sim1070/day$.

\hfill \break

\begin{figure}[htbp]
    \centering
    \includegraphics[width=0.99\textwidth]{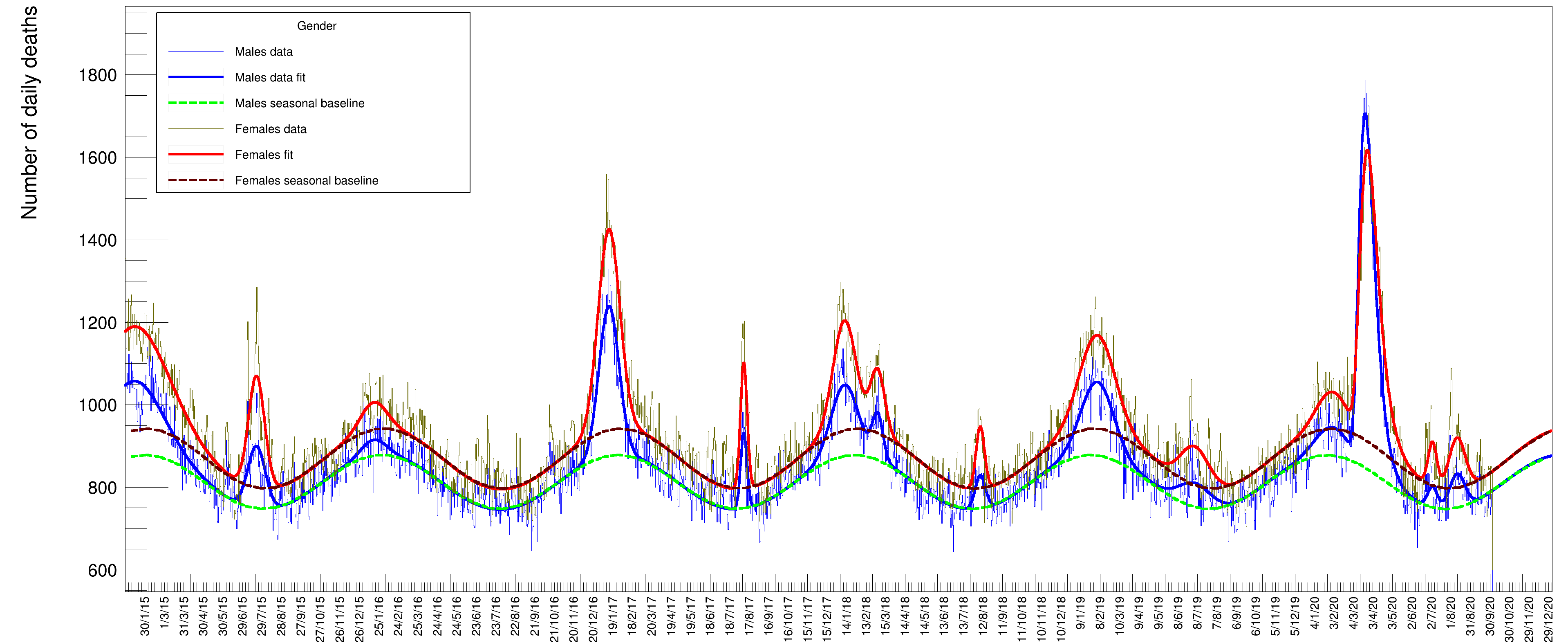}
    \caption{Number of daily casualties for males and females of all ages.}
    \label{fig:figAllGenders}
\end{figure} 

Another information that can be extracted from the data is the relative amount of deaths between genders. Fig~\ref{fig:figAllGenders} shows the distribution of males and females (summed over all ages) superimposed with the relative fits. In this case, since the two samples have a rather large statistical amount, both fits have been performed with all parameters free to vary. These numbers need to be corrected by the relative number of males and females in the Italian population. The fraction of males in 2020 was $48.7\%$ while females were $51.3\%$ \cite{reWeighted}: we compute a mortality factor (for each gender) by normalizing the yields to $29\,050\,096$ and $30\,591\,392$ (the respective number of males and females of the total Italian population by January $1^{st}$ 2020). 
The resulting values (multiplied by $100\,000$) are listed in Table~\ref{table:allGenders} under the columns \emph{Mortality}. While the absolute number of female deaths is higher than the males one in every year of the time series, the opposite seems true for the 2020 peak. After re-weighting this small discrepancy between genders this assertion remains basically true for all peaks except the 2020 one, where the mortality turns out to be larger for males than for females.

The fraction of casualties for the two genders turns out to be about the same, at the level of one standard deviation in all the years, till 2019 included. 

\begin{table}[htbp]
\centering
\begin{tabular}{|| c | c | c |c | c | c | c | c | c ||} 
 \hline
 \multicolumn{1}{||c| }{}                 &
 \multicolumn{4}{ |c| }{\textbf{Males}}   &
 \multicolumn{4}{ |c||}{\textbf{Females}} \\
 \hline
 $g_i$ & \emph{Yield} & \emph{Mortality} & \emph{Peak} & \emph{Duration}
 & \emph{Yield} & \emph{Mortality} & \emph{Peak} & \emph{Duration}\\ [0.5ex] 
 \hline\hline
  1 & 20\,279 $\pm$ 788 & 69.8 & 10/1/2015 & 178 & 32\,405 $\pm$ 929 & 105.9 & 10/1/2015 & 206 \\  
  2 &  5\,132 $\pm$ 258 & 17.7 & 21/7/2015 & 53  &  8\,401 $\pm$ 266 & 27.5  & 21/7/2015 & 49  \\  
  3 &  1\,734 $\pm$ 394 & 6.0  & 17/1/2016 & 76  &  3\,198 $\pm$ 441 & 10.5  & 17/1/2016 & 80  \\  
  4 & 13\,907 $\pm$ 357 & 47.9 & 13/1/2017 & 62  & 20\,281 $\pm$ 404 & 66.3  & 13/1/2017 & 66  \\  
  5 &  2\,252 $\pm$ 156 & 7.8  &  9/8/2017 & 19  &  3\,976 $\pm$ 174 & 13.0  &  9/8/2017 & 21  \\   
  6 &  8\,875 $\pm$ 468 & 30.6 & 11/1/2018 & 83  & 11\,830 $\pm$ 415 & 38.7  & 11/1/2018 & 71  \\   
  7 &  2\,192 $\pm$ 268 & 7.5  &  3/3/2018 & 33  &  4\,139 $\pm$ 305 & 13.5  &  3/3/2018 & 43  \\   
  8 &  1\,612 $\pm$ 194 & 5.5  &  8/8/2018 & 31  &  2\,473 $\pm$ 203 & 8.1   &  8/8/2018 & 27  \\   
  9 & 11\,066 $\pm$ 462 & 38.1 &  4/2/2019 & 99  & 15\,273 $\pm$ 509 & 49.9  &  4/2/2019 & 109 \\   
 10 &  3\,728 $\pm$ 341 & 12.8 &  4/7/2019 & 99  &  5\,224 $\pm$ 352 & 17.1  &  4/7/2019 & 87  \\    
 11 &  4\,112 $\pm$ 443 & 14.2 & 31/1/2020 & 99  &  5\,124 $\pm$ 439 & 16.7  & 31/1/2020 & 93  \\    
 12 &     964 $\pm$ 182 & 3.3  &  4/7/2020 & 29  &  1\,764 $\pm$ 182 & 5.8   &  4/7/2020 & 27  \\    
 13 &  2\,489 $\pm$ 236 & 8.6  & 11/8/2020 & 47  &  3\,826 $\pm$ 253 & 12.5  & 11/8/2020 & 49  \\    
 \hline
\end{tabular}
\caption{Results of the fit to the data set divided in a sample of males and another of females (of all ages) in Fig.~\ref{fig:figAllGenders}. The meaning of the \emph{Mortality} column is described in the text.}
\label{table:allGenders}
\end{table}
\begin{table}[htbp]
\centering
\begin{tabular}{|| c | c | c | c | c | c | c | c ||} 
 \hline
 \multicolumn{4}{ |c| }{\textbf{Males}}   &
 \multicolumn{4}{ |c||}{\textbf{Females}} \\
 \hline
 \emph{Yield} & \emph{Mortality} & \emph{Peak} & \emph{Duration}&
 \emph{Yield} & \emph{Mortality} & \emph{Peak} & \emph{Duration}\\ [0.5ex] 
 \hline\hline
 27\,240 $\pm$ 366 & 93.8 & 22/3/2020 & 63 & 26\,079 $\pm$ 395 & 85.2 & 26/3/2020 & 72 \\
 \hline
\end{tabular}
\caption{Results of the fit to the data set (for the Gompertz peak only) divided in two samples of males and  females (of all ages) in Fig.~\ref{fig:figAllGenders}. The meaning of the \emph{Mortality} column is described in the text.}
\label{table:allGendersGompertz}
\end{table}

\hfill\break

\section{Comparison between different data sets}
The data set provided by ISTAT \cite{ISTAT-Data} and used for the present analysis is not the only one publicly available: the Dipartimento della Protezione Civile (DPC) data set \cite{ProtezioneCivile-Data} provides a somewhat different kind of information regarding the number of deaths in the context of the COVID-19 pandemic. In particular, the data record, which begins February $24^{\mathrm{th}}$, 2020, contains the number of daily deaths directly attributed to the current pandemic, whereas the ISTAT one only refers to recorded deaths regardless of their cause. 

\begin{figure}[htbp]
    \centering
    \includegraphics[width=0.99\textwidth]{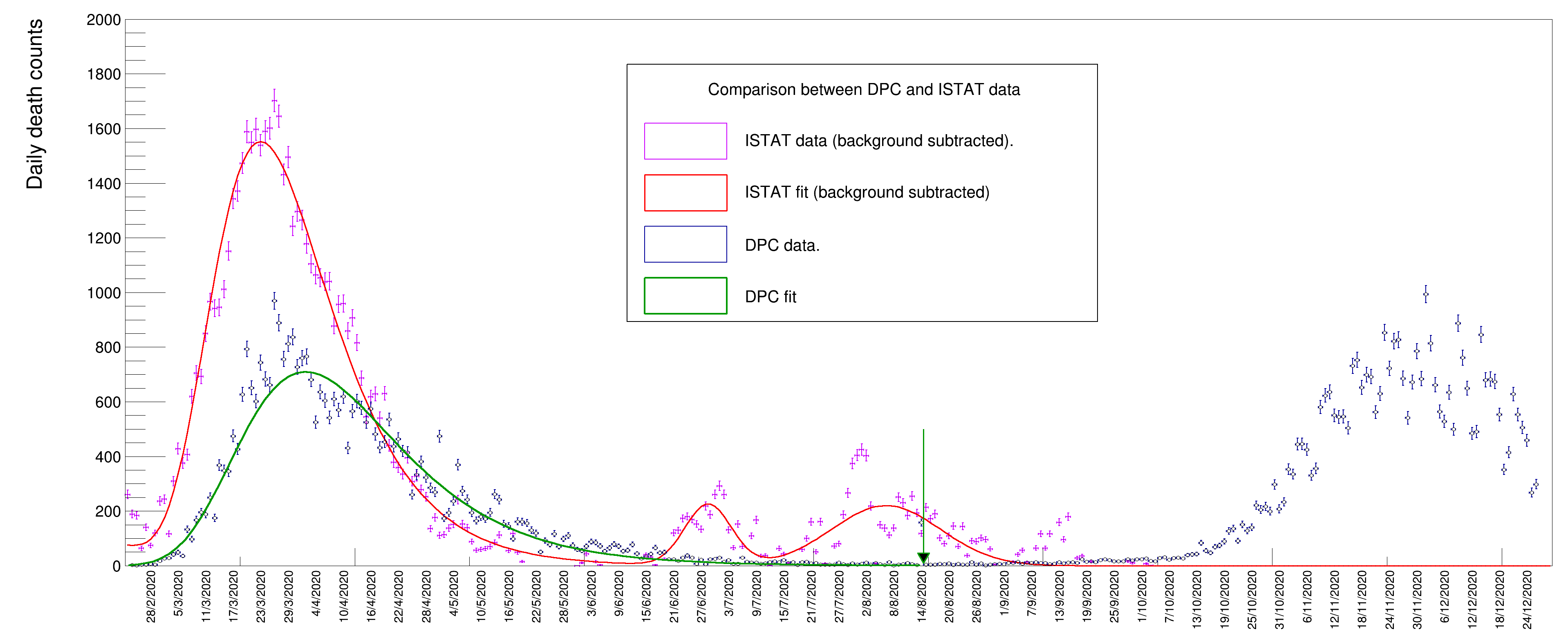}
    \caption{Comparison between ISTAT and DPC data samples}
    \label{fig:figComparison}
\end{figure}

A plot of the data from these two disparate sources is shown in Fig.~\ref{fig:figComparison}. The magenta points (and the accompanying fit result of a Gompertz derivative function in red) correspond to the ISTAT data sample: these data are a subset of those displayed in Fig.~\ref{fig:figAll}, specifically those between the dates of February $24^{\mathrm{th}}$ and September $30^{\mathrm{th}}$ 2020, with the entries in each bin replaced with the difference between the actual counts and the contribution due to the underlying wave. This subtraction of the background of the data allows for a direct comparison between the ISTAT and DPC data, the latter doesn't requires a subtraction procedure being unaffected by a background.

The DPC sample is shown as blue dots (with the corresponding Gompertz fit superimposed in green). A clear peak is visible around spring 2020 together with a second one during fall 2020, corresponding, respectively, to the first and the second wave of the 2020 pandemic. It is worthwhile to note that the DPC data reports the day when the death was finally registered, unlike the case of the ISTAT data, which records the actual day of death, thus introducing a potential delay of a few days between the two samples, visible as a translation of the green line with respect to the red one.

The DPC data shows a spike corresponding to August $15^{\mathrm{th}}$, due to the fact that a certain number of deaths were not correctly reported in the preceding weeks and were recovered assigning that day as the actual death date. In order to compare the yield returned by the fit to the value provided by the ISTAT data we had to exclude the contributions from the second pandemic peak: we decided to introduce a cutoff value while computing the sum of entries of the DPC sample in correspondence to August $16^{\mathrm{th}}$, a day when the minimum number of casualties was reached between the two pandemic waves, therefore including also the spike. The cutoff date is shown in Fig.~\ref{fig:figComparison} as a vertical green arrow.

The sum in that period (the blue dots) results to be $35\,468$. 

On the other hand, the yield obtained for the ISTAT sample is the one reported in Table~\ref{table:2}, namely $54\,387\pm557$, resulting from the integral of the Gompertz peak (the yields of two peaks at around July and August are therefore not included).
The difference in the number of deaths from these two samples amounts to $18\,919 \pm 557$. This strikingly large difference could be due to several different reasons, such as an excessive pressure on the Italian health system in the early stages of the pandemic which prevented a certain number of patients with diseases other than COVID-19 to be safely treated in hospitals and emergency rooms. We have no elements in the data that can allow us to discern the different contributions to this discrepancy and an exhaustive discussion about this outcome is beyond the scope of this article. 

\section{Additional considerations}
\begin{figure}[htbp]
    \centering
    \includegraphics[width=0.99\textwidth, height=0.25\paperheight]{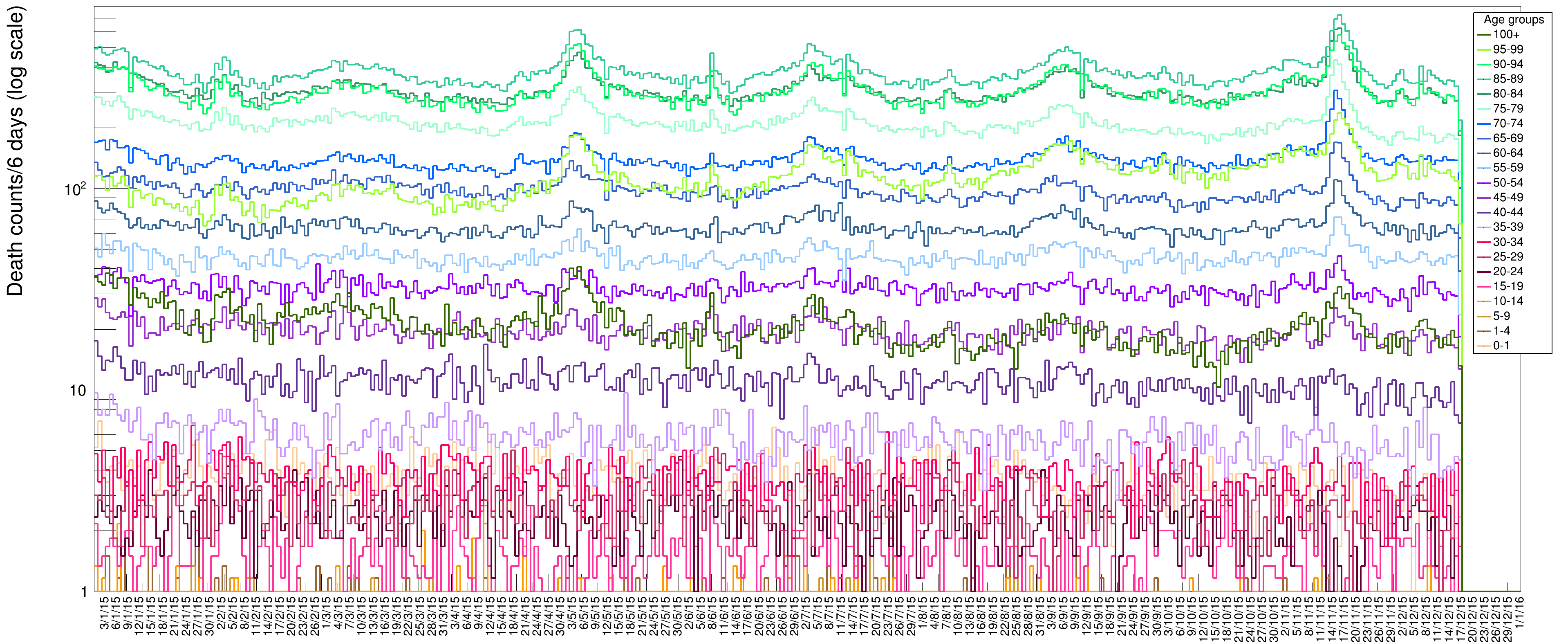}
    \caption{ISTAT data set with disentangled age categories. The data are binned in groups of six days each for an enhanced visualization clarity.}
    \label{fig:figDisentangleAll}
\end{figure}

\begin{figure}[htbp]
    \centering
    \includegraphics[width=1.05\textwidth, height=0.25\paperheight]{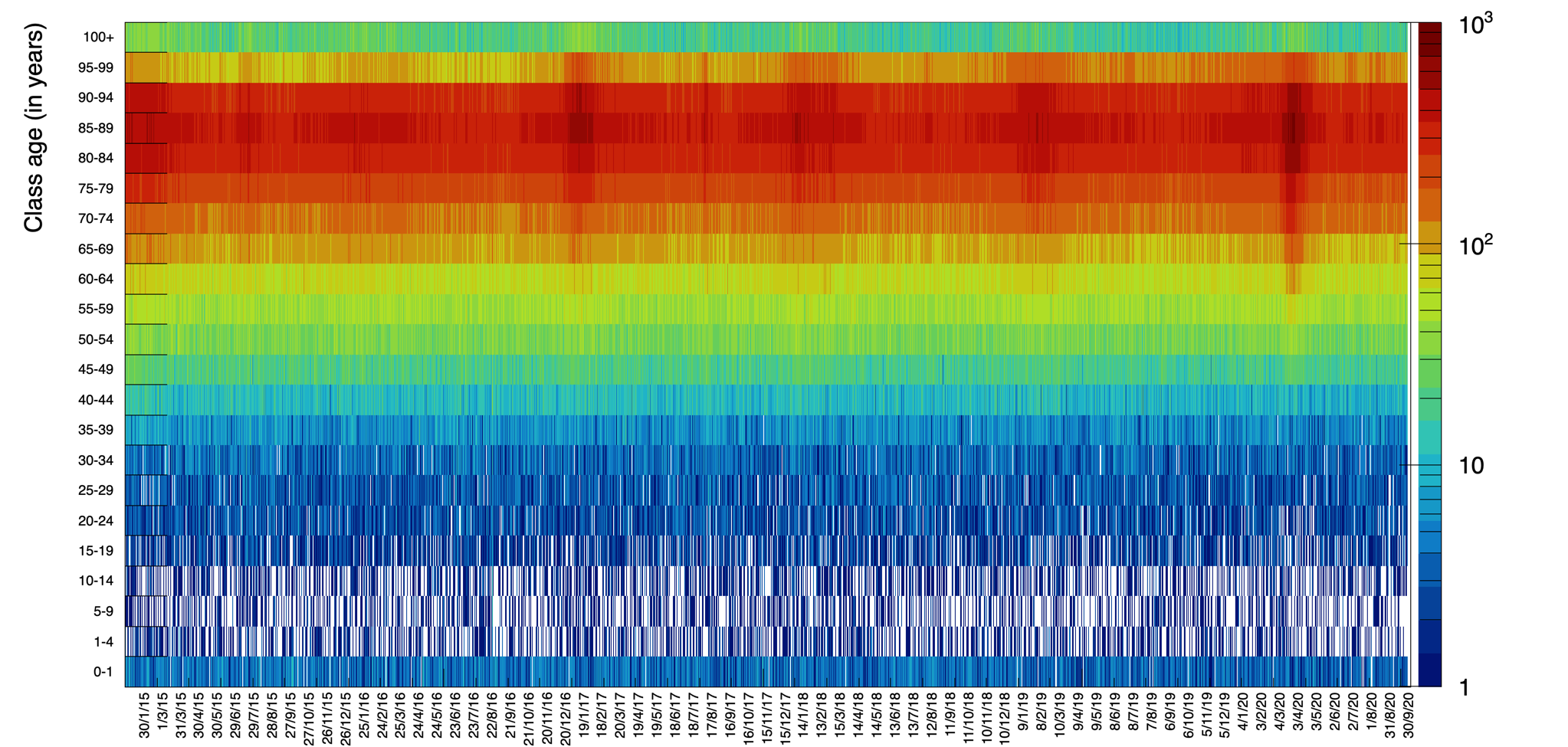}
    \caption{Scatter plot of the class age versus day of death. The dark red spots show the age and the date corresponding to highest values of deaths incidence.}
    \label{fig:figScatter}
\end{figure}

The rich data sample provided by ISTAT allows for various additional visualizations.
In Fig.~\ref{fig:figDisentangleAll} we display data for ages in groups of 4 years to visualize the increase of death probability with age: it becomes more evident what was already shown in Fig.~\ref{fig:figDisentangle}, namely the fact the young people tend to die with a rather flat probability along each year, while progressively higher age tend to suffer more from illnesses in specific seasonal periods. Each bin in this plot contains the number of deaths lumped together from six contiguous days.
In Fig.\ref{fig:figScatter} we present a scatter plot of death rates as a function of the day of the year (for the six years from 2015 to 2020) versus the age category. This graphical representation clearly illustrates the higher probability of death for the age category 70--95 with respect to the others.

\hfill\break

Each value in the ISTAT data sample comes with a geographical tagging marker, allowing for a categorization of the number of deaths in different parts of Italy.

\begin{figure}[htbp]
    \centering
    \includegraphics[width=0.99\textwidth, height=0.25\paperheight]{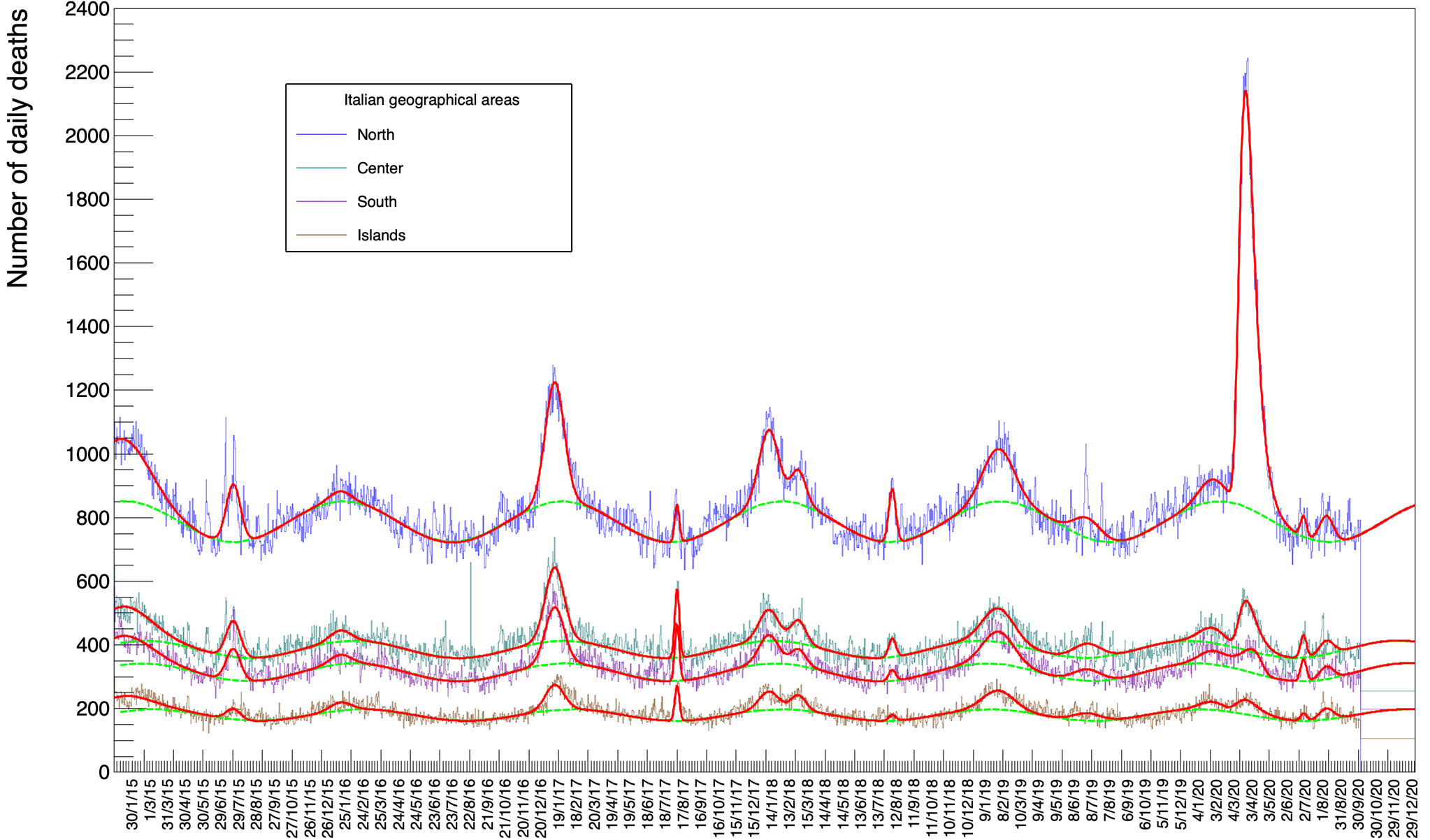}
    \caption{ISTAT data set with the Italian geographical areas disentangled (their definition is detailed in the text).}
    \label{fig:figAllRegions}
\end{figure}

Fig.~\ref{fig:figAllRegions} shows the fits for each of the four zones in Italy, namely \emph{North}, \emph{Center}, \emph{South} and \emph{Islands}\footnote{The subdivision is arbitrary and we have defined \emph{\textbf{North}} as the sum of values for the following regions: \emph{Piemonte, Valle d'Aosta, Liguria, Lombardia, Trentino-Alto Adige, Veneto, Friuli-Venezia Giulia, Emilia-Romagna}. \emph{\textbf{Center}} comprises \emph{Toscana, Umbria, Marche, Lazio, Abruzzo and Molise}, \emph{\textbf{South}} includes \emph{Campania, Puglia, Basilicata and Calabria}. Finally \emph{\textbf{Islands}} corresponds to \emph{Sicilia and Sardegna}}.
The fits in these plots correspond to minimization procedures with all parameters free.

Table \ref{table:12} reports the value of the Gaussian and Gompertz integral for these different regions. In order to compare these values between different zones, in Table \ref{table:13} we report the same values but normalized to the relative amount of registered inhabitants \cite{anagrafe}. 

A visual inspection of Fig.~\ref{fig:figAllRegions} shows the magnitude of the peak in the winter/spring of 2020 for the North of Italy which is not matched by a comparably populated peak for the Center, South and Islands.
Table \ref{table:13} confirms this impression numerically: while values of each column, for a given row (normalized by population), are comparable between zones, the value of the Gompertz peak in the North remains much bigger (actually by a factor from 10 to over 20).

\begin{table}[htbp]
\centering
\begin{tabular}{|| c | c | c | c | c ||} 
 \hline
 $g_i$ & \emph{North} & \emph{Central} & \emph{South} & \emph{Islands} \\ [0.5ex] 
 \hline\hline
  1 & 23\,046 $\pm$ 518  & 13\,293 $\pm$ 366  & 10\,670 $\pm$ 333 & 5\,730 $\pm$ 249 \\
  2 &  5\,520 $\pm$ 193  &  3\,409 $\pm$ 139  &  2\,995 $\pm$ 126 &    999 $\pm$  91 \\
  3 &  1\,333 $\pm$ 235  &  1\,501 $\pm$ 166  &  1\,250 $\pm$ 151 & 1\,120 $\pm$ 115 \\
  4 & 14\,924 $\pm$ 254  &  9\,184 $\pm$ 183  &  7\,058 $\pm$ 165 & 3\,135 $\pm$ 121 \\
  5 &  1\,143 $\pm$ 106  &  2\,078 $\pm$  85  &  1\,716 $\pm$  77 & 1\,066 $\pm$  58 \\
  6 &  9\,393 $\pm$ 246  &  4\,027 $\pm$ 170  &  3\,676 $\pm$ 156 & 2\,344 $\pm$ 119 \\
  7 &  3\,213 $\pm$ 203  &  2\,233 $\pm$ 144  &  1\,576 $\pm$ 129 & 1\,451 $\pm$ 102 \\
  8 &  2\,511 $\pm$ 135  &     876 $\pm$  93  &     511 $\pm$  82 &    274 $\pm$  61 \\
  9 & 10\,692 $\pm$ 302  &  6\,865 $\pm$ 213  &  6\,620 $\pm$ 197 & 3\,884 $\pm$ 150 \\
 10 &  3\,711 $\pm$ 255  &  2\,631 $\pm$ 180  &  2\,168 $\pm$ 161 & 1\,374 $\pm$ 122 \\
 11 &  3\,737 $\pm$ 264  &  2\,505 $\pm$ 186  &  2\,033 $\pm$ 171 & 1\,361 $\pm$ 130 \\
 12 &     836 $\pm$ 119  &     898 $\pm$  87  &     871 $\pm$  79 &    300 $\pm$  57 \\
 13 &  2\,477 $\pm$ 182  &  1\,286 $\pm$ 131  &  1\,188 $\pm$ 117 & 1\,026 $\pm$  90 \\
 \hline\hline
 Gompertz & 45\,350 $\pm$ 33   &  6\,063 $\pm$ 201  &  3\,559 $\pm$ 205 & 2\,158 $\pm$ 150 \\
 \hline
\end{tabular}
\caption{Integral of the various peaks of Fig.\ref{fig:figAllRegions} detailed for the 4 Italian geographical areas defined in the text}.
\label{table:12}
\end{table}
\begin{table}[htbp]
\centering
\begin{tabular}{|| c | c | c | c | c ||} 
 \hline
 $g_i$    &  \emph{North (\%)}    & \emph{Central (\%)}  & \emph{South (\%)}    & \emph{Islands (\%)} \\ [0.5ex] 
 \hline\hline
 1       & 0.0835 $\pm$ 0.0019 & 0.0990 $\pm$ 0.0027 & 0.0881 $\pm$ 0.0027 & 0.0883 $\pm$ 0.0038 \\
 2       & 0.0200 $\pm$ 0.0007 & 0.0254 $\pm$ 0.0010 & 0.0247 $\pm$ 0.0010 & 0.0154 $\pm$ 0.0014 \\
 3       & 0.0048 $\pm$ 0.0009 & 0.0112 $\pm$ 0.0012 & 0.0103 $\pm$ 0.0012 & 0.0173 $\pm$ 0.0018 \\
 4       & 0.0540 $\pm$ 0.0009 & 0.0684 $\pm$ 0.0014 & 0.0583 $\pm$ 0.0014 & 0.0483 $\pm$ 0.0019 \\
 5       & 0.0041 $\pm$ 0.0004 & 0.0155 $\pm$ 0.0006 & 0.0142 $\pm$ 0.0006 & 0.0164 $\pm$ 0.0009 \\
 6       & 0.0340 $\pm$ 0.0009 & 0.0300 $\pm$ 0.0013 & 0.0303 $\pm$ 0.0013 & 0.0361 $\pm$ 0.0018 \\
 7       & 0.0116 $\pm$ 0.0007 & 0.0166 $\pm$ 0.0011 & 0.0130 $\pm$ 0.0011 & 0.0224 $\pm$ 0.0016 \\
 8       & 0.0091 $\pm$ 0.0005 & 0.0065 $\pm$ 0.0007 & 0.0042 $\pm$ 0.0007 & 0.0042 $\pm$ 0.0009 \\
 9       & 0.0387 $\pm$ 0.0011 & 0.0511 $\pm$ 0.0016 & 0.0547 $\pm$ 0.0016 & 0.0599 $\pm$ 0.0023 \\
10       & 0.0134 $\pm$ 0.0009 & 0.0196 $\pm$ 0.0013 & 0.0179 $\pm$ 0.0013 & 0.0212 $\pm$ 0.0019 \\
11       & 0.0135 $\pm$ 0.0010 & 0.0187 $\pm$ 0.0014 & 0.0168 $\pm$ 0.0014 & 0.0210 $\pm$ 0.0020 \\
12       & 0.0030 $\pm$ 0.0004 & 0.0067 $\pm$ 0.0006 & 0.0072 $\pm$ 0.0007 & 0.0046 $\pm$ 0.0009 \\
13       & 0.0090 $\pm$ 0.0007 & 0.0096 $\pm$ 0.0010 & 0.0098 $\pm$ 0.0010 & 0.0158 $\pm$ 0.0014 \\
 \hline\hline                       
Gompertz & 0.164  $\pm$ 0.001  & 0.045  $\pm$ 0.001  &  0.029 $\pm$ 0.002  &  0.033 $\pm$ 0.002  \\
 \hline
\end{tabular}
\caption{Mortality in the four Italian zones: the quoted values are obtained by normalizing the values of Table \ref{table:12} to the number of inhabitants in those same regions taken from \cite{anagrafe}, corresponding to the population registered at December $31^{st}$, 2020.}
\label{table:13}
\end{table}
\break\hfill
\section{Conclusions}
The data provided by ISTAT allow for a detailed quantitative estimate of the number of deaths excesses with respect to a baseline. This baseline is represented by a sinusoidal variation of the number of deaths which turns out to be almost perfectly in phase with the yearly seasonal cycle. We presented a study of these excesses evaluated by a statistical interpolation of the data based on a $\chi^2$ minimization method using a function which is the sum of a sinusoidal wave, a number of Gaussian distributions to represent the excesses above the sinusoid and, finally, a Gompertz derivative to model the asymmetric peak of spring 2020.
The overall fit resulted satisfactory in terms of the final $\chi^2$ and pull distributions, describing the 2014 data points with just 46 parameters. This allows for a quantitative definition of the properties of all the peaks, along with a precise determination of the errors.
In this study we discussed the methodology adopted for the interpolations and analyze different samples by disentangling genders, ages and locations. A comparison has also been carried out between the number of deaths provided by ISTAT in the period corresponding to the first wave of the pandemic and the numbers provided by DPC in the same period for the deaths directly attributed to COVID-19. We found a rather large discrepancy, amounting to $18919 \pm 557$ deaths over a total of $54387 \pm 557$.
We have no elements in the data that can allow us to discern the different contributions to this discrepancy and an exhaustive discussion about it is beyond the scope of this article. 

As a final remark, we think this study once more underlines the importance of a unified protocol of data collection and the online availability of these same data under a sheared Open Data international agreement. Open Data repositories with useful data already exist (ISTAT and DPC are good examples) but they are not exhaustive in the number of information provided. Other repositories, containing valuable data for improved analyses are usually restricted or not compliant with the FAIR~\cite{FAIR} approach, one of the prerequisites of the Open Data paradigm. These shortcomings hamper the possibility of further in-depth studies of the pandemics effects and its evolution by a large number of scholars. INFN is very active in this field and has recently implemented an Open Access/Open Data repository\cite{OAR}, containing also, among many other documents and data sets, the whole ensemble of results produced by our group.

\section{Acknowledgements}
The present work has been done in the context of the INFN CovidStat project that produces an analysis of the public Italian COVID-19 data. The results of the analysis are published and updated on the website 
{\it \link{https://covid19.infn.it}}. We are grateful to Mauro Albani, Marco Battaglini, Gianni Corsetti and Sabina Prati of ISTAT for useful insights and discussions. We wish also to thank Daniele Del Re and Paolo Meridiani for useful discussions. The project has been supported in various ways by a number of people from different INFN Units. In particular, we wish to thank, in alphabetic order: Stefano Antonelli (CNAF), Fabio Bredo (Padova Unit), Luca Carbone (Milano-Bicocca Unit), Francesca Cuicchio (Communication Office), Mauro Dinardo (Milano-Bicocca Unit), Paolo Dini (Milano-Bicocca Unit), Rosario Esposito (Naples Unit), Stefano Longo (CNAF), and Stefano Zani (CNAF). We also wish to thank Prof. Domenico Ursino (Universit\`a Politecnica delle Marche) for his supportive contribution.

\end{document}